\documentclass[12pt]{iopart}
\usepackage{iopams}
\usepackage{color,multicol}
\usepackage{graphicx}
\usepackage{pstricks}
\usepackage{epsfig}

\begin{document}

\title{Bandstructure meets many-body theory: The LDA+DMFT method}
\author{K. Held$^{1}$, O.\ K.\ Andersen$^{1}$, M.
Feldbacher$^{1}$, A. Yamasaki$^{1\dagger }$, and Y.-F.\ Yang$^{1,2}$}

\begin{abstract}
\textit{Ab initio} calculation of the electronic properties of materials is
a major challenge for solid state theory. Whereas the experience of forty
years has proven density functional theory (DFT) in a suitable, \textit{e.g.}
local approximation (LDA) to give a satisfactory description in case
electronic correlations are weak, materials with strongly correlated, say $d$%
- or $f$-electrons remain a challenge. Such materials often exhibit
\textquotedblleft colossal\textquotedblright\ responses to small changes of
external parameters such as pressure, temperature, and magnetic field, and
are therefore most interesting for technical applications.

Encouraged by the success of dynamical mean field theory (DMFT) in dealing
with \textit{model} Hamiltonians for strongly correlated electron systems,
physicists from the bandstructure and many-body communities have joined
forces and have developed a combined LDA+DMFT method for treating materials
with strongly correlated electrons \textit{ab initio}. As a function of
increasing Coulomb correlations, this new approach yields a weakly
correlated metal, a strongly correlated metal, or a Mott insulator.

In this paper, we introduce the LDA+DMFT by means of an example, LaMnO$_{3}$%
. Results for this material, including the \textquotedblleft
colossal\textquotedblright\ magnetoresistance of doped manganites are
presented. We also discuss advantages and disadvantages of the LDA+DMFT
approach.
\end{abstract}


\address{$^1$ Max-Planck Institut f\"ur Festk\"orperforschung,  D-70569
Stuttgart, Germany} 
\address{$^2$  Department of Physics, University of California, Davis, California 95616, USA
} \ead{k.held@fkf.mpg.de}

\submitto{\JPCM}


\section{Introduction}

The challenges of solid-state theory are to qualitatively understand
material's properties and to calculate these, quantitatively and reliably.
This task is particularly difficult if electronic correlations are as strong
as they are in many materials containing transition and rare-earth elements.
Here, the Coulomb interactions between the valence electrons in $d$- and $f$%
-orbitals can be strong. The reason for this difficulty is that the standard
approach, the \emph{local} density approximation (LDA) \cite{DFT}, for
calculating material's properties relies on the electronic correlations in
jellium, a weakly correlated system. For more correlated materials, the
electronic density is strongly varying and the assumption of a constant
density for treating exchange and correlation is not warranted. That is, the
exact functional of density-functional theory \cite{DFT}, which -if known-
would allow the treatment of correlated materials, is certainly \emph{%
non-local}. Another  difficulty is the
construction of  functionals 
beyond ground-state properties, e.g., for spectral properties.

\begin{figure}[t]
\epsfig{file=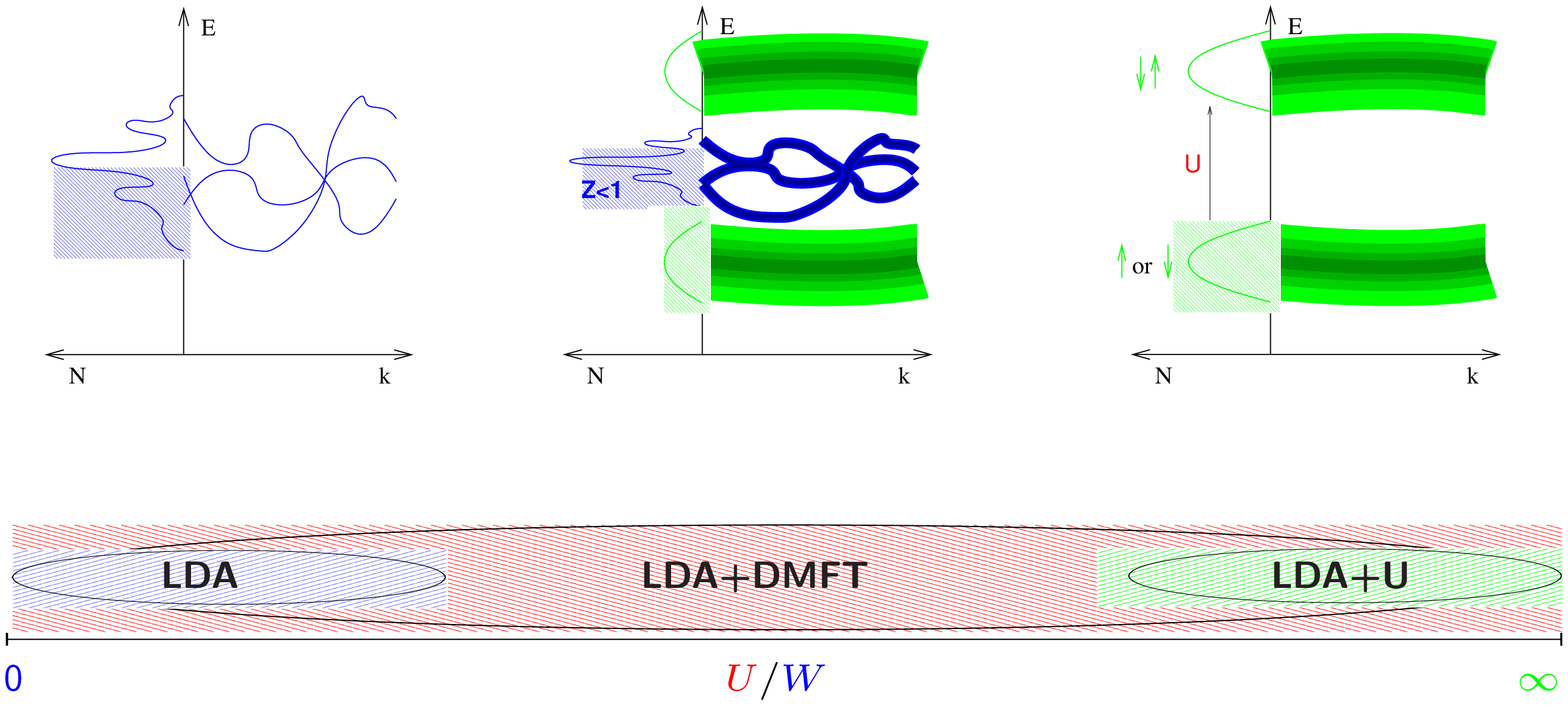,width=12.4cm,angle=90} \vspace{-5.2cm}
\caption{ With increasing Coulomb interaction $U$ (relative to the bandwidth 
$W$), we go from a weakly correlated metal via a strongly correlated metal
with renormalized quasiparticles to a Mott insulator with a gap in the
spectrum. The LDA bandstructure correctly describes the weakly correlated
metal; LDA+U does so, with some restrictions \protect\cite{Sangiovanni05},
for the Mott insulator; and LDA+DMFT gives the correct physics in the entire
parameter regime. (reproduced from \protect\cite{LDADMFTreview2}) }
\label{Fig:LDALDADMFTLDAU}
\end{figure}

In this situation, we have seen a break-through brought about by a new
method, LDA+DMFT \cite{LDADMFT,LDADMFTreview1,LDADMFTreview2}, which merges
LDA with dynamical mean-field theory (DMFT) \cite{DMFT1,DMFT2,DMFTreview} to
account for the electronic correlations. This approach has been developed in
an effort by theoreticians from the bandstructure and the many-body
communities joining two of the most successful approaches of their
respective community. By now, LDA+DMFT has been successfully employed to
calculate spectral, transport, and thermodynamic properties of various
transition-metal oxides, magnetic transition metals, and rare-earth metals
such as Ce and Pu; see \cite{LDADMFTreview1,LDADMFTreview2} for reviews.
Depending on the strength of the Coulomb interaction, LDA+DMFT gives a
weakly correlated metal as in LDA, a strongly correlated metal, or an
insulator as illustrated in Fig.\ \ref{Fig:LDALDADMFTLDAU}.

In the following Section, we will introduce this method by the example of a
particular material currently of immense interest: the
colossal-magnetoresistance (CMR) material, LaMnO$_{3}$. Results for the
parent compound \cite{LaMnO3LDADMFT}, as well as for doped manganites \cite%
{CMRLDADMFT} are presented in Section \ref{Sec:results}. Finally, in Section %
\ref{Sec:conclusion}, we conclude and discuss the pros and cons of LDA+DMFT.

\section{LDA+DMFT in a nutshell}

\label{Sec:LDADMFT}

\begin{figure}[t]
\vspace{.4cm}
\par
\centerline{\bf LDA+DMFT in a nutshell} \vspace{.5cm}
\par
\noindent{\bf 1)} \textcolor{blue}{\bf LDA} calculation $%
\Longrightarrow$ $\textcolor{blue}{\epsilon^{\mathrm{LDA}}_{\mathbf{k} lm}}$ 
\vspace{-.3cm}
\par
\includegraphics[clip=false,width=11.25cm]{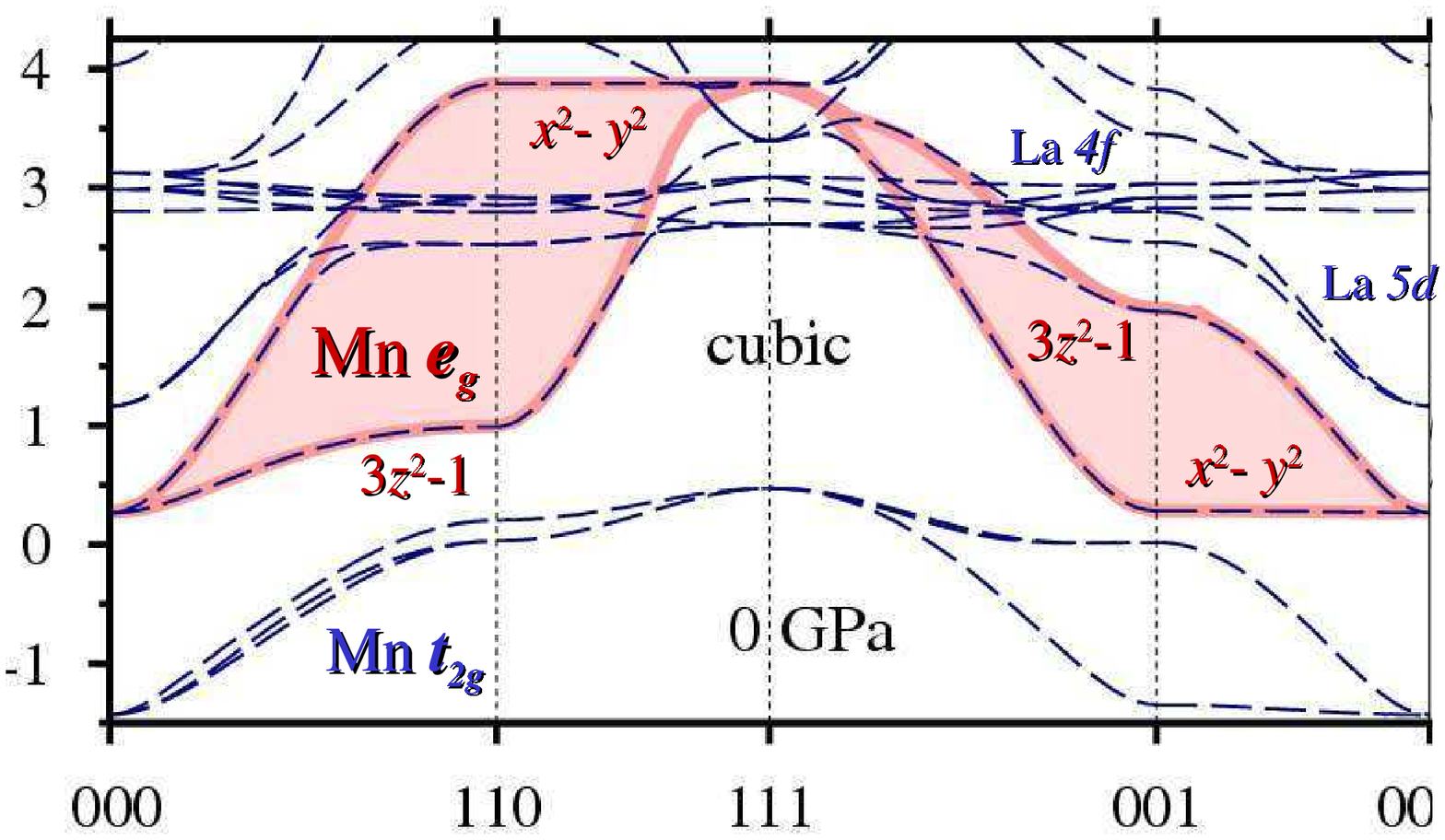} \vspace{-.002cm}
\par
\noindent 
\fcolorbox[rgb]{0.,0.,0.}{1.,1.,0.8}{\parbox{15.cm}{\parbox{15.cm}{
{\bf 2)} Supplement \textcolor{red}{$\epsilon^{\rm LDA}_{{\bf k} lm}$} by local \textcolor{red}{Coulomb interactions}
\vspace{-.6cm}

 \begin{eqnarray} 
\hat{H}&=&  \sum_{l,m=1}^{2} \sum_{{\bf k} \sigma}
  {\hat{c}}^{\dagger}_{{\bf k} l\sigma} \textcolor{blue}{\epsilon^{\rm LDA}_{{\bf k} lm}}  {\hat{c}}^{\phantom{\dagger}}_{{\bf k} m\sigma} 
   - 2  \textcolor{red}{\cal J} \;     \sum_{m i \sigma\tilde{\sigma}} {\hat{c}}^{\dagger}_{{i m\sigma}} {{\bf  \tau}}_{\sigma \tilde{\sigma}}   {\hat{c}}^{\phantom{\dagger}}_{{i} m \tilde{\sigma}} \;
   {\hat{\bf  S}}^{t_{2g}}_{i}\nonumber \\&&
    \!\!+    \textcolor{red}{ U}   \sum_{m i} 
    \hat{n}_{im\uparrow}\hat{n}_{im\downarrow}
     + \sum_{i \,  \sigma \tilde{\sigma}} 
    (  \textcolor{red}{U'}\!-\!\delta_{\sigma \tilde{\sigma}} \textcolor{red}{J})  \;
    \hat{n}_{i  1 \sigma} \hat{n}_{i 2  \tilde{\sigma}}
\label{Eq:LDADMFT}
 \end{eqnarray} 
}}} \vspace{.5cm}
\par
\noindent%
\fcolorbox[rgb]{0,0,0}{1,1.,.8}{\parbox[b]{15.0cm}{

\noindent {\bf 3)}  Solve    $\hat{H}$ by  DMFT 
\vspace{.4cm}

\includegraphics[clip=false,width=15cm]{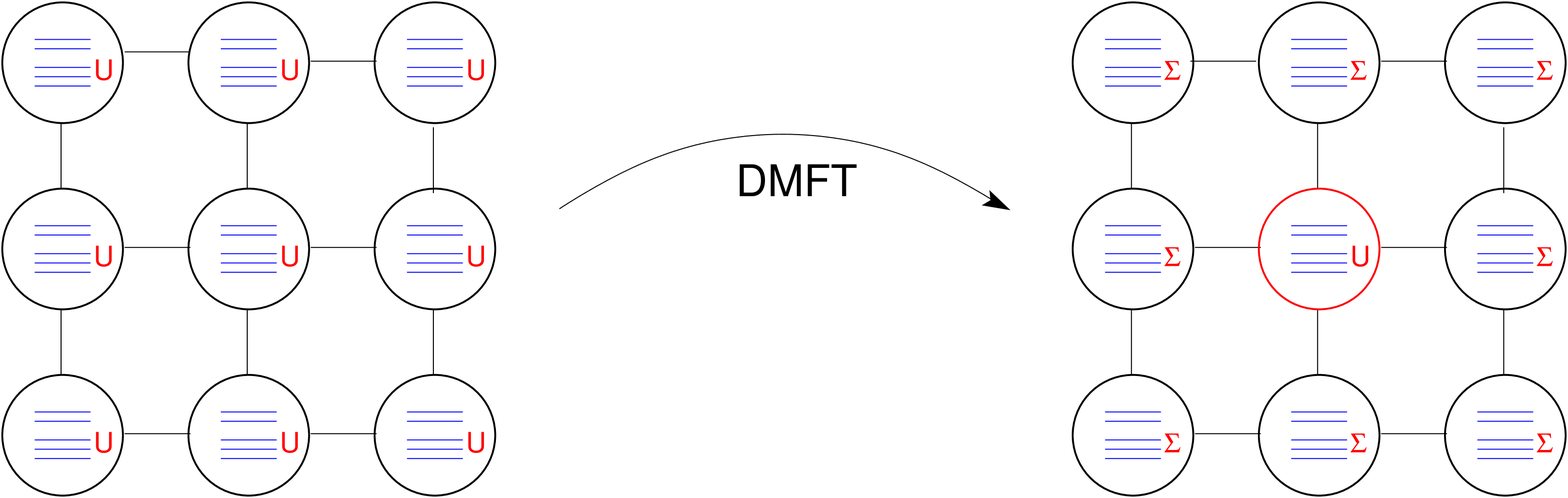}

\vspace{.2cm}
material specific lattice problem $\hat{\cal H}$
\hfill Anderson impurity problem\phantom{aaaaa}
 
\hfill + Dyson eq.\phantom{aaaaaaasaaaaa}
}} \vspace{-18.25cm}
\par
\hspace{11.2cm} 
\fcolorbox[rgb]{1.,1.,1.}{1.,1.,1.}{\parbox{7.cm}{\parbox{6.8cm}{
\includegraphics[clip=false,width=4.7cm]{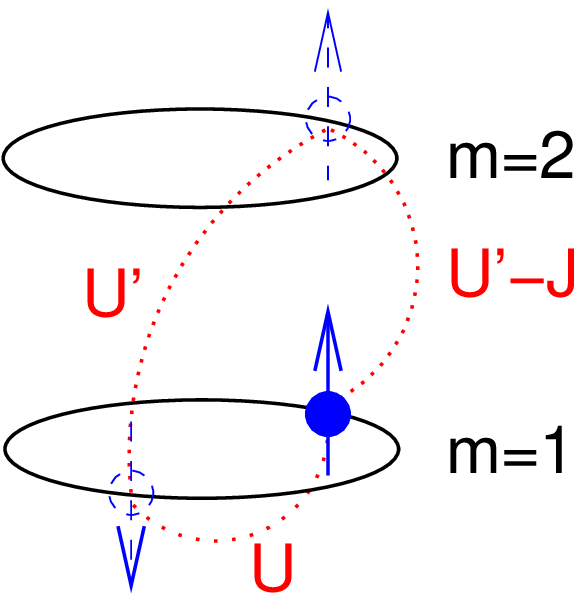}
\vspace{.5cm}

\hspace{1.2099cm}  \textcolor{blue}{\LARGE $\uparrow$}
\vspace{-.2cm}
 
\hspace{1.3cm}  \textcolor{blue}{\LARGE $|$}

\vspace{-1.1cm} 

\hspace{1.25cm} \textcolor{blue}{\Large $\bullet$}
\vspace{-.3cm} 

\hspace{1.25cm} \textcolor{blue}{\Large $\bullet$}
\vspace{-.3cm} 

\hspace{1.25cm} \textcolor{blue}{\Large $\bullet$}
\vspace{-2.cm} 

\hspace{3cm}  \textcolor{red}{\Huge )}
 
\vspace{-.7cm} 

\hspace{3.7cm}  \textcolor{red}{\Large $\cal J$}
}}} \vspace{14.57cm}
\caption{The three steps of an LDA+DMFT calculation.}
\label{Fig:LDADMFT}
\end{figure}

The first step of an LDA+DMFT calculation is the calculation of the LDA
bandstructure. This paramagnetic bandstructure for our LaMnO$_{3}$ example
is shown at the top of Fig.\ \ref{Fig:LDADMFT} for the ideal cubic
structure. We employed the $N$th order muffin-tin orbital (NMTO) basis set 
\cite{NMTO}. As we will later restrict the electronic correlations to the
strongly interacting, more localized $d$-and $f$-orbitals, we need to
identify these orbitals in the LDA calculation. In the case of LaMnO$_{3}$
where each Mn$^{3+}$ ion is in the nearly cubic environment at the centre of
an oxygen octahedron, these are the three lower-lying $t_{2g}$ and the two
higher lying $e_{g}$ (3$d$) orbitals. Since Mn$^{3+}$ has the $d^{4}$
configuration, the first three $d$ electrons occupy the $t_{2g}$ orbitals
forming a spin $3/2$ according to Hund's rule. This leaves us with one
electron per Mn in the two $e_{g}$ orbitals. With a $t_{2g}^{\uparrow
\uparrow \uparrow }e_{g}^{\uparrow }$ mean-field occupation (LSDA or LDA+U),
only the $e_{g}^{\uparrow }$-like LDA bands would cross the Fermi level. For
transition-metal oxides, one typically -in present day LDA+DMFT
calculations- restricts the DMFT calculation to the low-energy bands
crossing the Fermi level. Here, we employ NMTO downfolding \cite{NMTO} for
obtaining the effective LDA Hamiltonian for two Mn $e_{g}$ orbitals,
labeled $m=1$ and $2$ in Fig.\ \ref{Fig:LDADMFT}. This Hamiltonian \cite%
{CMRLDADMFT} can be written in terms of the $2\times 2$ orbital matrix ${%
\epsilon _{\mathbf{k}lm}^{\mathrm{LDA}}}$ whose diagonalization gives the
LDA bandstructure, see first term of Eq.\ (\ref{Eq:LDADMFT}). In other
calculations, e.g., for Ce \cite{Ce}, all ($spdf$) valence orbitals have
been taken into account. As shown in the top part of Fig.\ \ref{Fig:LDADMFT}%
, the downfolded Hamiltonian (red bands) describes the LDA bandstructure of
the LaMnO$_{3}$ $e_{g}$ orbitals very well. If other basis sets, such as
plane waves, are used, the construction of a minimal set of well localized
orbitals can be more involved. But this is also possible, e.g.,\ through
Wannier-function projection \cite{WFP,WFP2}.

The second step of an LDA+DMFT calculation is to supplement the LDA
Hamiltonian by the local Coulomb interaction which is responsible for the
electronic correlations, see second part of Fig.\ \ref{Fig:LDADMFT}. In
general, the Coulomb interactions can be expressed, e.g., by Racah
parameters \cite{Racah}. In actual calculations however, this term has been hitherto 
restricted to the inter-orbital Coulomb interaction $U^{\prime }$
and Hund's exchange $J$ (there is also a pair hopping term of the same
size), see Fig.\ \ref{Fig:LDADMFT} top part, right hand side. The
intra-orbital Coulomb interaction $U=U^{\prime }+2J$ follows by symmetry.
For a parameter-free (\textit{ab initio}) calculation, the (screened)
Coulomb interactions have to be determined. As LDA+DMFT calculations for a
prototype transition-metal oxide, SrVO$_{3}$ \cite{SrVO3}, and a prototype
rare-earth metal, Ce \cite{Ce}, showed, such  \textit{ab initio} LDA+DMFT
calculations employing the constrained LDA \cite{cLDA} for determining $%
U^{\prime }$ and $J$ work very well. There is some uncertainty of $\sim 0.5\;
$eV \cite{LaTiO3} in $U^{\prime }$ due to the ambiguity in defining the $d$
orbitals, leading to an additional error besides the
 LDA and DMFT approximations involved.
 This can be a problem if one is close to a transition and hence
sensitive to small changes of $U^{\prime }$, as is e.g.\ the case for V$%
_{2}$O$_{3}$ \cite{V2O3} which is close to a Mott-Hubbard transition. 
But usually results do not
alter dramatically upon changing  
$U^{\prime }$ by $\sim 0.5\;$eV \cite{LaTiO3}.

In the
case of LaMnO$_{3}$, the half-occupied $t_{2g}$ orbitals prevent us from
using standard constrained LDA calculations. Hence, we took the $U^{\prime }=3.5\,$eV
value from the literature \cite{Park} and $J=0.75\,$eV from the spin-up/spin-down
splitting of a ferromagnetic LSDA calculation. In the following, the three $%
t_{2g}$ electrons are taken into account as a (classical) spin-$3/2$,
coupled through Hund's exchange $\mathcal{J}$ to the $e_{g}$ spin, see the
second term of the Hamiltonian (\ref{Eq:LDADMFT}) in Fig.\ \ref{Fig:LDADMFT}.

The third step of the LDA+DMFT calculation is to employ DMFT for solving the
many-body Hamiltonian (\ref{Eq:LDADMFT}). Had we used the unrestricted
Hartree-Fock (static mean-field) approximation instead, we would have the
LDA+$U$ approach \cite{LDAU}. In DMFT, we replace the Coulomb interaction on
all sites, but one, by a self-energy. Electrons interact on this single site
and still move through the whole lattice. However, on the other sites, they
propagate through the medium given by the self-energy instead of the
interaction. This is the DMFT approximation, which hence neglects non-local
vertex contributions. The emerging DMFT single-site problem is equivalent to
an auxiliary Anderson impurity model \cite{DMFT2}
which has to be solved self-consistently together
with the standard relation (Dyson eq.) between self energy and Green function.
DMFT becomes exact \cite{DMFT1} if the number of neighbors $Z\rightarrow
\infty $ and is a good approximation for a three dimensional system with
many neighboring lattice sites. In particular, it provides for an accurate
description of the major contribution of electronic correlation: the local
correlations between two $d$- or $f$-electrons on the same site. For more
details on DMFT, see \cite{DMFTreview,LDADMFTreview1,LDADMFTreview2}.

If the electron density changes after the DMFT calculation, we have to go
back to the first step and recalculate the LDA Hamiltonian for this new
density. In contrast to the frequency-dependent spectral function, the
electron density itself however only changes to a lesser degree. This
self-consistency is therefore often left out.

An alternative point of view, besides the above-mentioned Hamiltonian one,
is the spectral density-functional theory \cite{SDFT}. This theory states
that the ground state energy $E[\rho (\mathbf{r}),G_{ii}(\omega )]$ is a
functional which depends not only on the electron density $\rho (\mathbf{r}),
$ but also on the local Green function (spectral function) $G_{ii}(\omega )$%
. LDA+DMFT is an approximation to this, in principle exact functional, in
the same spirit as the LDA is to the exact density functional.

\section{Results for manganites}

\label{Sec:results} 
\begin{figure}[b]
\centerline{\includegraphics[width=.8\columnwidth]{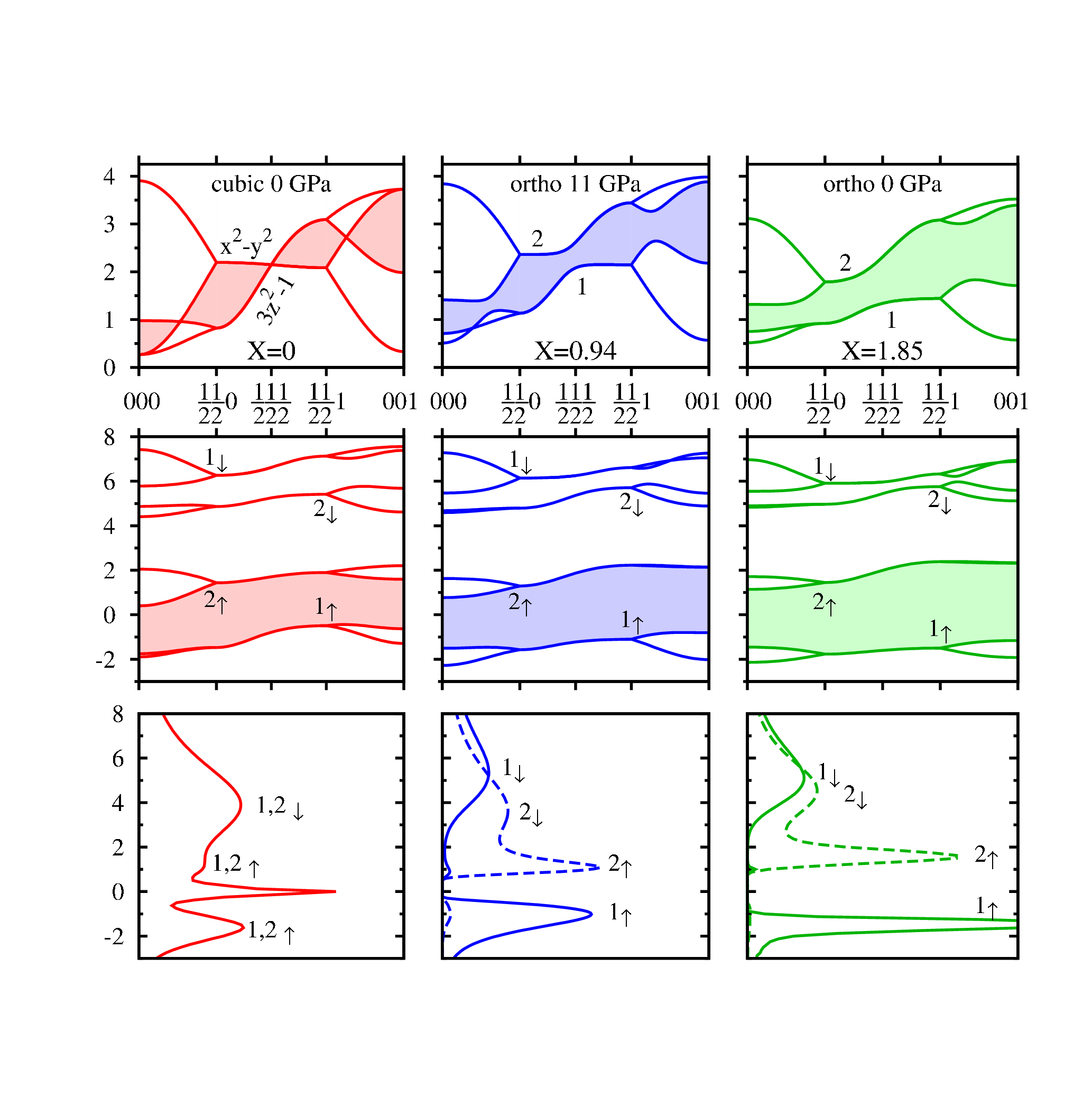}}
\caption{ Bandstructure of paramagnetic LaMnO$_{3}$ as obtained in LDA
(top), LDA+U (center), and LDA+DMFT (bottom; $k$-integrated spectrum) for
the experimental (orthorhombic) crystal structure at 0 GPa (right), 11GPa
(center) and an artificial cubic structure with the same volume as at 0GPa
(left). Energies are in  eV with the Fermi energy being 0;
the unit of $\mathbf k$ is $\pi$.
 For the correct insulating behavior and size of the gap, both
Coulomb correlations and the crystal field splitting due to the orthorhombic
distortion are needed (bottom right). For more details see \protect\cite%
{LaMnO3LDADMFT}, from which this figure was reproduced.}
\label{Fig:Bands}
\end{figure}

Let us now turn to the results obtained for LaMnO$_{3}$.
Fig.\ \ref{Fig:Bands} compares LDA, LDA+$U$ and LDA+DMFT results for the
real Jahn-Teller- and GdFeO$_{3}$-distorted, orthorhombic crystal structure
at 0~GPa and 11~GPa, as well as for an artificial cubic structure with the
same volume as at 0~GPa. 
All LDA Hamiltonians were calculated for the paramagnetic phase, which is
the stable one at 300~K. Even though the lattice distortion leads to a
crystal-field splitting of the two $e_{g}$ bands in the LDA, these bands
still overlap. Hence, without electronic correlations, \emph{i.e.} without $%
U^{\prime }$, the plain vanilla LDA predicts a metal; it cannot describe the
insulating paramagnet observed experimentally. If we now consider the
many-body Hamiltonian (\ref{Eq:LDADMFT}) and treat it in the unrestricted
Hartree-Fock approximation, we obtain the LDA+$U$ bands shown in the middle
panel of Fig.\ \ref{Fig:Bands}. We see that the crystal-field splitting
becomes largely enhanced with the result that LaMnO$_{3}$ becomes an
insulator, even in the cubic phase at compressions exceeding those for which
the material is experimentally known to be metallic \cite{Loa}. However,
such effects are overestimated in the LDA+$U$ approximation. We therefore
turn to LDA+DMFT which does a better job in this respect. In the cubic
phase, and even at normal pressure, LDA+DMFT yields metallic behavior.
Hence, both Coulomb interaction and crystal-field splitting are necessary
and work hand in hand to make LaMnO$_{3}$ insulating at normal pressure. The
resulting gap is slightly smaller than 2~eV as in experiment. Our LDA+DMFT
calculations show that for LaMnO$_{3}$ to be metallic at pressures above the
experimental 32 GPa, some distortion must persist. For further details, see 
\cite{LaMnO3LDADMFT}.

Most fascinating, both from the point of view of basic physics and of
materials engineering, is the \textquotedblleft colossal\textquotedblright\
magnetoresistance \cite{CMR} of doped manganites such as La$_{1-x}$Sr$_{x}$%
MnO$_{3}$. At low temperatures, doped manganites are bad-metallic
ferromagnets, whereas at high temperatures, they are insulating \cite%
{LaMnO3,Okimoto95a} for a wide range of doping. Since Sr dopes holes in LaMnO%
$_{3}$, one would generally expect a metallic behavior. As the lattice
distortion fades away upon doping, we can start from a cubic crystal
structure, for which the nearest-neighbor tight-binding hopping gives
already an accurate description of the LDA bandstructure, as shown in \cite%
{LaMnO3LDADMFT}. Even without the static lattice distortion, we must however
include the distortion in the form of phonons. We do so by the two
Jahn-Teller phonons coupled to the $e_{g}$ electrons through the
electron-phonon coupling constant $g$.

\begin{figure}[tbp]
\noindent \includegraphics[clip=true,width=4.5cm,angle=270]{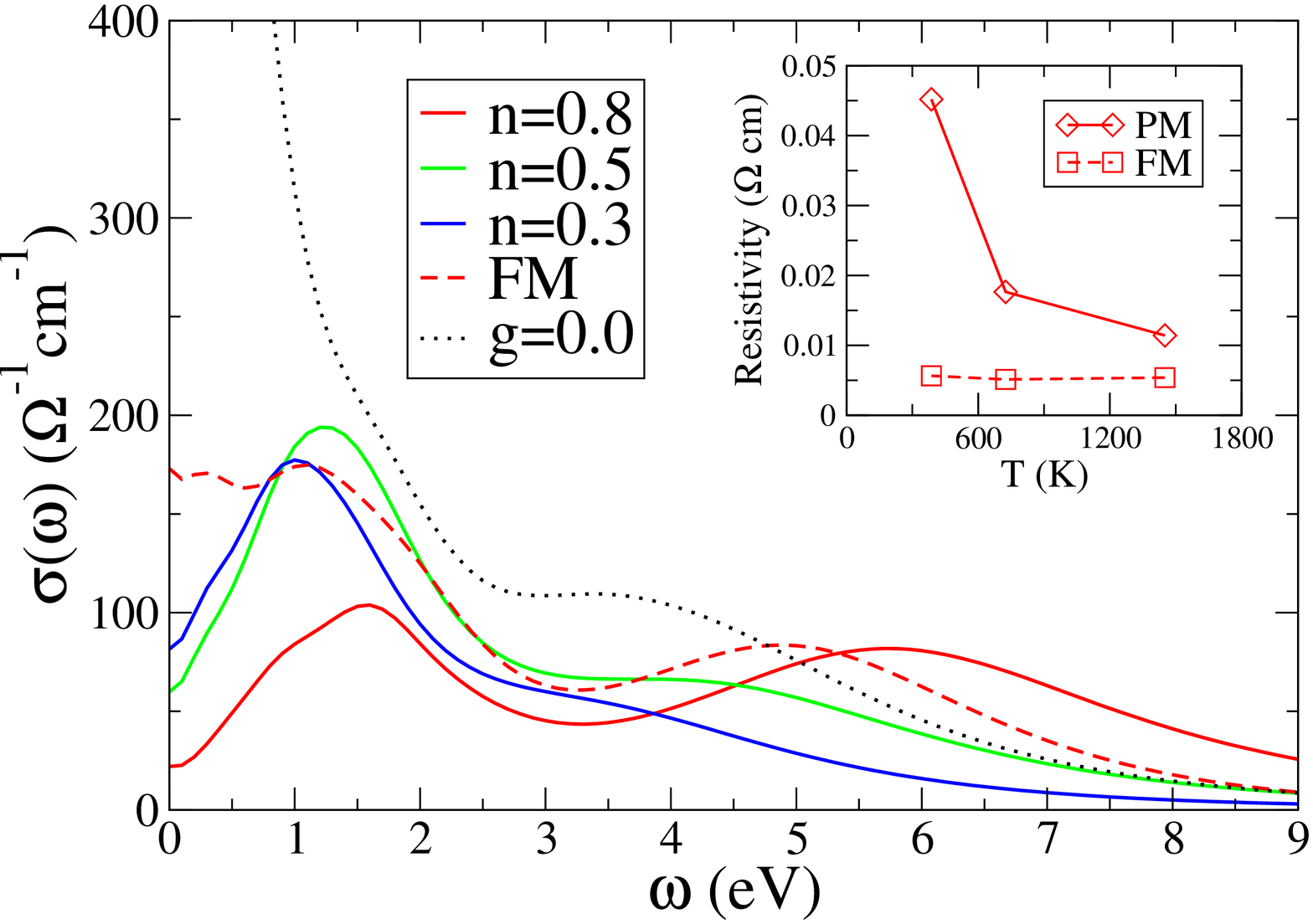} 
\vspace{-5.4cm}
\par
\hfill \includegraphics[clip=false,width=7.cm]{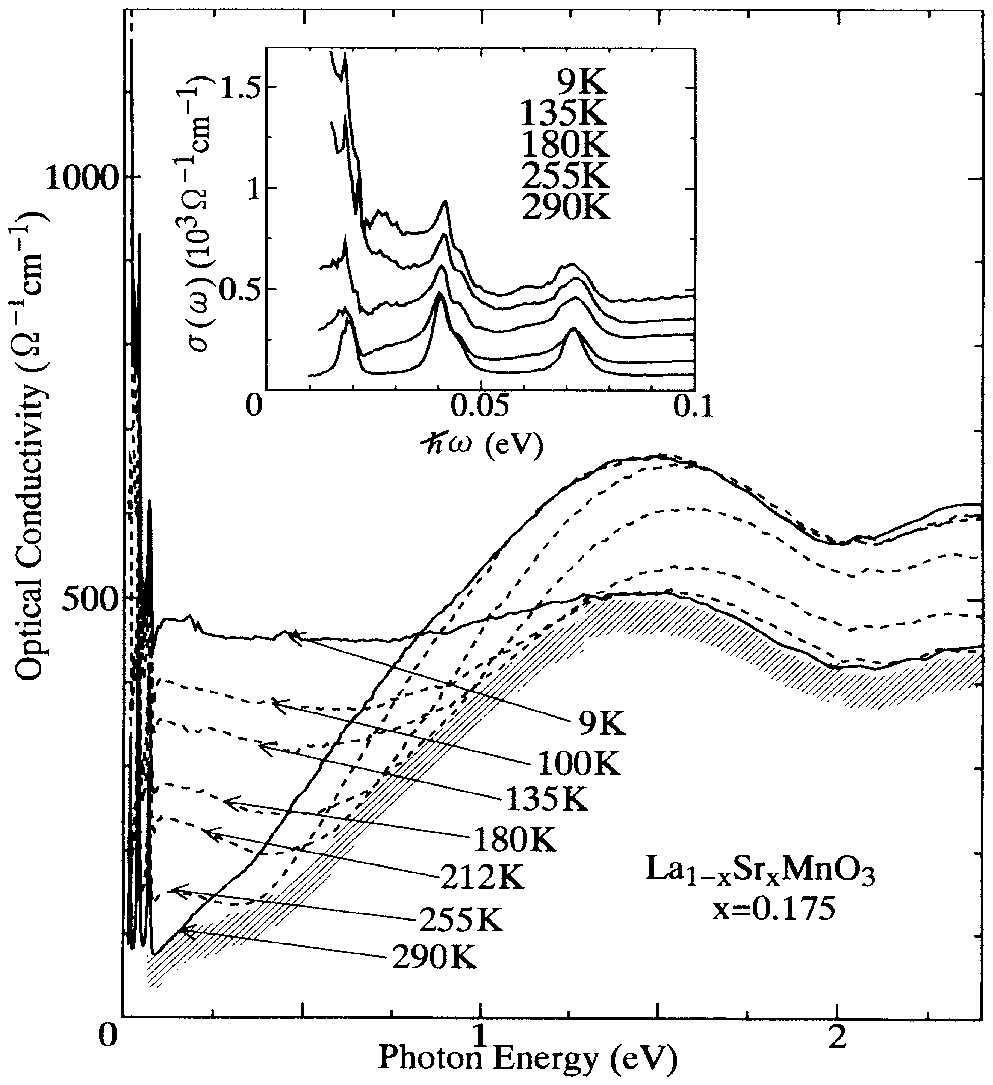}
\caption{ LDA+DMFT optical conductivity (left; reproduced from \protect\cite%
{CMRLDADMFT}) in comparison with experiment (right; reproduced from 
\protect\cite{Okimoto95a}) for the paramagnetic (PM) phase of La$_{1-x}$Sr$%
_x $MnO$_3$ ($n=1-x$ electrons/site). The dotted line shows a metallic Drude
peak in the absence of electron-phonon coupling; the dashed line the `bad'
metallic behavior in the ferromagnetic phase (FM) at $x=0.2$.\newline
Inset: In the PM (FM) phase, the resistivity shows insulating (`bad'
metallic) behavior so that the application of a magnetic field results in a
``colossal'' magnetoresistance.\label{Fig:CMR} }
\label{Fig:LaMnO3}
\end{figure}

These local Holstein phonons are described by a single frequency $\omega $.
Our DMFT calculation, see \cite{CMRLDADMFT} for details, shows again that
Coulomb interaction and electron-phonon coupling mutually support each
other: On lattice sites with a single electron the Jahn-Teller coupling
leads to a (dynamic) splitting of the two $e_{g}$ levels which is strongly
enhanced by the Coulomb interaction. In this way the electrons are localized
as a lattice polaron, explaining the unusual experimental properties of
doped manganites \cite{polaronAM}. Fig. \ref{Fig:CMR} shows as an example
the optical conductivity and the resistivity as a function of temperature.
As one can see, the paramagnet has a (pseudo-)gap at low frequencies and is
therefore insulating-like. In contrast the ferromagnet is a (bad) metal.
Since the ferromagnetic phase can be stabilized by a small magnetic field, a
\textquotedblleft colossal\textquotedblright\ magnetoresistance emerges.

\section{Conclusion and perspectives}

\label{Sec:conclusion} Using the example of LaMnO$_3$, we introduced the LDA+DMFT
approach and presented some of the results obtained. Let us conclude this
paper, by outlining the advantages and disadvantages of LDA+DMFT. The most
striking advantages are:

\begin{enumerate}
\item We can now calculate electronic properties of \emph{strongly
correlated 3d- and 4f-materials} with an accuracy comparable to that of the
LDA for electronically weakly correlated materials.

\item As the name \emph{dynamical} mean-field theory suggests, the \emph{%
dynamics} of the electrons is included, as are the excited states. One
always calculates the excitation spectrum. These states are effective-mass
renormalizations of the LDA one-particle states. Actually, we even have 
\emph{two} effective-mass renormalizations of the LDA dispersion relation $%
\epsilon _{k}$ and a kink in-between \cite{kink}. Also finite life times due
to the electron-electron scattering and metal-insulator transitions are
included.

\item Besides the spectral function for the addition or removal of single
electrons, also correlation functions can be calculated. From these two
functions, all physical quantities can be calculated:  spectra,
transport properties,  thermodynamics. All this
naturally arises from a well-defined theory without the need to construct,
e.g., from the LDA an effective Heisenberg model and from this,
susceptibilities and critical temperatures (see e.\ g.\ \cite{HeldManganites,FeNi}
for such DMFT calculations) 
\end{enumerate}

\noindent With so many advantages, there are also disadvantages:

\begin{enumerate}
\item While the DMFT includes the major part of the electronic correlations,
i.e., the \emph{local} correlations induced by the local Coulomb
interaction, non-local correlations are neglected. These give rise to
additional, interesting physics, typically at lower temperatures, e.g.,
magnons, quantum criticality, and possibly high-temperature
superconductivity. Recently, cluster \cite{cDMFT} and diagrammatic
extensions \cite{DGA} of DMFT have been developed to overcome this obstacle.

\item Another drawback is the computational cost for solving the Anderson
impurity model. The numerical effort of the standard quantum Monte Carlo
(QMC) simulations grows as $M^{2}(1/T)^{3}$ with a big prefactor for the
Monte-Carlo statistics. Here, $M$ is the number of interacting orbitals.
When $n$ inequivalent ions with $d$- or $f$-orbitals are included in a
supercell, the effort grows linearly $\sim n^{1}$. This means that typical
LDA+DMFT calculations at room temperature require some hours on present day
computers. The biggest problem is the $1/T^{3}$ increase of the
computational effort. However, more recently developed QMC approaches, such
as projective QMC \cite{PQMC} and continuous-time QMC \cite{ctQMC} at least
mitigate this drawback.

\item Presently, the most important point preventing the widespread
application of LDA+DMFT in academia and in industry is the lack of
standardized program packages. But the inclusion of DMFT into well spread
LDA codes, such as, e.g., the Vienna Ab initio Simulation Package (VASP) 
\cite{VASP}, will certainly be done in the near future.

\item A more principle disadvantage is the need to identify the interacting $%
d$- or $f$-orbitals. This is cumbersome if one starts with plane waves and
the result will also depend ---to some extent-- on the LDA basis set employed
and the procedure to define the $d$- or $f$-orbitals from these basis
function, e.g., via NMTO partial-wave downfolding or via Wannier-function
projection.
\end{enumerate}

With the pros clearly outweighing the cons, many of which have been or will
be mitigated, LDA+DMFT or variants such as GW+DMFT will be used more and
more in the future for calculations of correlated materials. With LDA and DMFT,
bandstructure has finally met many-body theory. The next step is to meet
industry.

\ack We would like to commemorate our young, dedicated coworker, A. Yamasaki, who
prominently contributed to the original work presented in this article.

\section*{References}

 $\dagger $ Deceased

\end{document}